\documentclass[doublecol]{epl2}
\usepackage{amsfonts}
\usepackage{graphics}
\usepackage{amsfonts}
\usepackage{amssymb}
\usepackage{amsthm}
\usepackage{newlfont}
\usepackage{amssymb,amsmath,latexsym}
\newcommand\beq{\begin{equation}}
\newcommand\eeq{\end{equation}}

\newcommand\bea{\begin{eqnarray}}
\newcommand\eea{\end{eqnarray}}

\newcommand\bi{\begin{itemize}}
\newcommand\ei{\end{itemize}}

\newcommand\non{\nonumber}

\renewcommand\hm{\mathcal H}

\newcommand\odd{1{\textendash}D}

\newcommand\ie{{\it{i.e.}}}

\newcommand\tdeg{{\textsf{2DEG~}}}

\newcommand\ir{{\textsf{IR~}}}

\newcommand\fqhld{{\textsf{FQHL}}}

\newcommand\fqhl{{\textsf{FQHL~}}}

\newcommand\rgd{{\textsf{RG}}}

\newcommand\rg{{\textsf{RG~}}}

\newif\ifboo \boofalse
\def\Review#1{\boofalse{\it #1},}

\def\Vol#1{\ifboo Vol. {\bf #1}\else{\bf #1}\fi}
\def\Year#1{\ifboo #1\else(#1)\fi}

\def\Page#1{\ifboo {\rm p. #1}\else{\rm #1}\fi}

\newcommand{\eps}{\varepsilon}

\title{Quantum Pump for Fractional Charge}

\author{Sourin Das \and Vadim Shpitalnik}
\shortauthor{Sourin Das \etal}

\institute{{Department of Condensed Matter Physics, Weizmann
Institute of Science, Rehovot 76 100, Israel\\
E-mail:
Sourin.Das@weizmann.ac.il;~Vadim.Shpitalnik@weizmann.ac.il}}

\abstract{We propose a theoretical scenario for pumping of
fractionally charged quasi-particle in the context of $\nu=1/3$
fractional quantum Hall liquid. We consider quasi-particle pumping
across an anti-dot level tuned close to the resonance. Fractional
charge pumping is achieved by slow and periodic modulation of
coupling of the anti-dot level to left and right moving edges of a
Hall bar set-up. This is attained by periodically modulating the
gate voltages controlling the couplings. In order to obtain
quantization of pumped charge in the unit of the electronic charge
fraction ($\nu e$) per pumping cycle in the adiabatic limit, we
argue that the only possibility is to tune the quasi-particle
operator to be irrelevant from being relevant in the
renormalization group sense, which can be accomplished by invoking
quantum Hall line junctions into the Hall bar geometry. We also
comment on possibility for experimental realization of the above
scenario.}

\begin{document}

\maketitle

\section{Introduction}
\begin{figure*}[htb]
\begin{center}
\includegraphics[width=12cm,height=4.5cm]{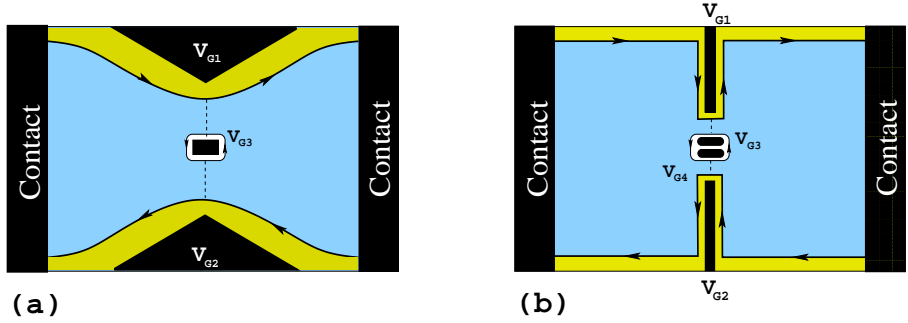}
\caption{(a) The conventional Hall bar geometry with an anti-dot
in the center produced by a top gate, $V_{G3}$.  (b) The proposed
geometry with two half line junctions invoked with an anti-dot in
the center which is produced by two top gates, $V_{G3}$ and
$V_{G4}$.
The dotted line shows that there is tunnelling from the
anti-dot to the left and right moving edges. The region containing
quantum Hall liquid is shown in blue colour.
 } \label{geometry}
\end{center}
\end{figure*}

There has been a considerable interest in the direction of
obtaining quantization of charge, pumped per cycle of pumping, in
units of charge associated with the fundamental excitation of the
quantum liquid in which the pump is operating. This issue was
first addressed in a classic article by Thouless~\cite{thouless}
followed by a series of
theoretical~\cite{hekking,aleiner,levinson,das2005,oreg,das2007sd}
and experimental works~\cite{kou,pothier,switkes}. Much of the
research activity in this direction till date has been carried out
in the context of Fermi liquids, which leads to quantization of
pumped charge in units of the electronic charge $e$. Also in case
of non-Fermi liquids, only pumping of electrons has been
considered~\cite{chamon,sen,gefen,novikov1,novikov2}. However the
case of non-Fermi liquids with fractionally charged excitations
has not been discussed much in the literature. An excellent
candidate of direct relevance in this regard is the fractional
quantum Hall liquid (\fqhld). A very interesting scenario in this
context was first proposed by Simon~\cite{simon} where quantum
charge pumping was considered across an anti-dot geometry in a two
terminal Hall bar set-up for the case of $\nu=1,1/3$ and other
abelian fractions of quantum Hall liquid. Charge pumping was
achieved by periodic modulation of gate voltages which would push
the left and the right moving edges of the Hall bar close to the
anti-dot hence resulting in periodic modulation of tunnel-coupling
of the left and right moving edges with the resonant level of the
anti-dot. Here the left and right moving edges act as
quasi-particle reservoirs for the quantum pump. As far as
quantization of pumped charge in unit of $\nu e$ in a single
pumping cycle for $\nu =1/m$ is concerned, this set-up has a
serious short-coming. This is because, the quantization of pumped
charge requires that the pumping contour in the parameter plane of
couplings of the anti-dot to the left and right moving edges must
enclose a quasi-particle
resonance~\cite{levinson,das2005,das2007sd} such that the
conductance on the pumping contour remains vanishingly small. This
amounts to saying that, for obtaining quantization in the
adiabatic limit the enclosed quasi-particle resonance should be
very sharp. On the contrary, we must keep in mind that the
quasi-particle tunnelling operator is a relevant
operator~\cite{kf_moon} and hence in the limit of low temperature,
zero bias and vanishingly small pumping frequency (\ir limit) it
will get renormalized and flow to the strong coupling limit
destroying the quasi-particle resonance completely. In this
letter, we propose a possible way to get around this hurdle. Our
idea is to convert the quasi-particle operator from being relevant
to an irrelevant (or marginal) one in the sense of renormalization
group (\rgd) flow which will in turn result in sharpening of the
quasi-particle resonance as we go to the \ir limit. Theoretically,
we show that this can be achieved by introducing strong, non-local
inter-edge repulsive interactions within the left and right moving
branches separately. To realize this within an experimentally
feasible set-up, we propose to invoke two line
junctions~\cite{kfline,fradkin} constructed from the left and
right moving branches, which operate as quasi-particle reservoirs
for the quantum pump as shown in Fig.~\ref{geometry}(b). Then,
this line junction geometry naturally imbibes the possibility for
realizing gate controlled inter-edge interactions leading to an
irrelevant (in the sense of \rg flow) quasi-particle tunnelling
operator in the strong interaction limit.

In this letter we show that, it is possible to obtain an exact
expression for pumped charge for a specific value of inter edge
interaction of the line junction as the effective tunnelling from
left to right edge through the anti-dot turns out to be marginal.
Then, using weak interaction \rg analysis~\cite{matveev}, we find
that the pumped charge gets quantized when the quasi-particle
tunnelling operator is irrelevant and the quantization is
destroyed, when tunnelling operator becomes relevant.
%
\section{Proposed device and its theoretical modelling}
The proposed set-up comprises of an anti-dot, tunnel-coupled to
two (half) line junctions acting as quasi-particle reservoirs (see
Fig.~\ref{geometry}(b)). We propose that the line junctions should
be fabricated by {\textsl{(i)}} very thin etching of the \tdeg
along the line perpendicular to the left and right moving edges
from the two sides of the Hall bar such that the etching ends in
the vicinity of the anti-dot region and then {\textsl{(ii)}}
depositing top gates at the etched regions. The fine etching
results in splitting of the \fqhl in two parts with very closely
spaced counter propagating edge states interacting only via
Coulomb interaction. The etching of the \tdeg prevents tunnelling
of electrons within the counter propagating edges of the line
junction. In addition, by negatively biasing the top gates,
$V_{G1}$ and $V_{G2}$, the effective distance between the two
counter propagating edges can be tuned resulting in tunability of
strength of Coulomb interaction between the two edges. We also
propose that the anti-dot should be produced by applying negative
bias to, not just one, but two top gates ($V_{G3}$ and $V_{G4}$ in
Fig.~\ref{geometry}(b)) so that the tunnel coupling of the
anti-dot with the line junction on its left and right side can be
tuned independently.

We model the edge states forming the line junction using chiral
Luttinger liquid theory~\cite{wen}. We assume the inter-edge
repulsive interaction between the counter propagating edges in each
line junction is faithfully described by a screened  Coulomb
interaction as the long range part is expected to be screened by
gates $V_{G1}$ and $V_{G2}$, on top of the line junction. We also
consider the anti-dot to be small enough so that the energy gap
between two consecutive quasi-particle resonant levels, $\Delta E$
is larger than all other energy scales in the problem. Hence one can
safely neglect the existence of all other levels in the anti-dot
except for the one which is tuned to be the closest to the Fermi
level in the lead and the anti-dot can be modelled as a single
resonant level. In this limit, the quasi-particle creation and
annihilation operators ($c,c^\dagger$) represent a hard core
anyon~\cite{averin} as is clarified below. Even though in general
the quasi-particle operator should obey exchange rules corresponding
to the fractional statistics associated with the quasi-particle, for
a single resonant level such statistical exchanges are irrelevant
owing to the fact that there are no possibilities for exchange of
quasi-particle in the set-up considered. Hence for all practical
purposes, $c$ and $c^\dagger$ can be treated as fermionic operators
as their true statistics is irrelevant in the set-up considered and
$\langle c^\dagger c \rangle$ can be either zero or one. The full
Hamiltonian for our system can therefore be written as
\begin{gather}
\hm = \hm_R+\hm_L+\hm_{tunn}+\hm_{ad}\cr = \sum_{i=R,L} \Bigg[
\frac{\pi v}{\hbar \nu} \int_{-L/2}^{L/2} dx \left[
(\rho_{i}(x))^2 + 2 \lambda \rho_{i}(x) \rho_{i} (-x)
 \right] + \cr
 [\Gamma_i (t)c^\dagger \psi_{i}(0) + h.c. ] \Bigg] + \eps (c^\dagger c)
\nonumber
\end{gather}
Henceforth, we set $\hbar=1$. Here, $\eps$ represents the
energy of resonant level with respect to the Fermi-level in the
leads and $\rho_{R/L}(x)$ represents the electronic density of the
right (left) moving branch and  can be expressed in terms of the
bosonic field as $\rho_{R/L}(x)=\pm (1/2\pi)
\partial_x \phi_{R/L}(x)$, which satisfies the commutation
relation, $[\phi_{i}(x),\phi_{i}(x')]=\pm i \pi \nu
~{\mathsf{Sgn}}(x-x')$. $c$ and $c^\dagger$ are the quasi-particle
annihilation and creation operators in the anti-dot level which
satisfy $\{c,c^\dagger\} = 1$ for the hard core anyon limit as
discussed earlier. $\psi_{R/L}(x)$ is the quasi-particle
annihilation operator in the line junction which is related to the
bosonic fields via the standard bosonisation identity,
$\psi_{R/L}(x)=(f_{R/L}/{(2 \pi \epsilon)^{\nu/2}}) e^{i
\phi_{R/L}(x)}$ where $\epsilon$ is the short distance cut-off and
$f_{R/L}$ is the Klein factor associated with the quasi-particle
operator. Henceforth we shall drop the Klein factors as
quasi-particle exchange processes are not relevant for transport
through a single resonant level \cite{rao}. $\Gamma_{R/L}(t)$ is
the time-dependent tunnel coupling and $\lambda$ is the strength
of screened Coulomb interaction between the counter propagating
edges of the line junctions.
 We now define new fields
\begin{gather}
\varphi_{R/L}=\frac{\phi_{R/L}^+ + \phi_{R/L}^-}{2}~; \quad
\vartheta_{R/L}=\frac{\phi_{R/L}^+ - \phi_{R/L}^-}{2} \nonumber
\end{gather}
where $\phi_{R/L}^{\pm}=\phi_{R/L}(\pm x)$ and the $\vartheta$
field satisfies the boundary condition, $\vartheta(x=0)=0$. These
new fields diagonalize the line junction Hamiltonian to give
\begin{gather}
\hm_{R/L}= \frac{v}{8  \pi \nu} \int_{0}^{L/2} dx \left[ k
(\partial_x \varphi_{R/L})^2 + \frac{1}{k} (\partial_x
\vartheta_{R/L})^2
 \right]
 \nonumber
\end{gather}
where $v = v_0\sqrt{{1-\lambda^2}}$ is the renormalized velocity
and $k = \sqrt{({1-\lambda})/({1+\lambda})}$ is the Luttinger
liquid parameter for $\nu=1$ case. Note that $0 \le k < 1$ for
repulsive interaction ($\lambda > 0$), while for attractive
interaction ($\lambda < 0$), $k > 1$. Now we re-scale the
$\varphi$ and $\vartheta$ fields as
\begin{gather}
\tilde \varphi_{R/L}=\sqrt{\frac{k}{\nu}} \, \varphi_{R/L}~; \quad
\tilde \vartheta_{R/L}= \frac{1}{\sqrt{k\nu}} \, \vartheta_{R/L}
\nonumber
\end{gather}
 to obtain
\begin{gather}
\hm_{R/L}= \frac{1}{8  \pi} \int_{0}^{L/2} dx \left[ (\partial_x
\tilde\varphi_{R/L})^2 +  (\partial_x \tilde\vartheta_{R/L})^2
 \right]
\label{five}
\end{gather}
and
\begin{gather}
\hm_{tunn} =  \sum_{i=R,L}\frac{1}{(2\pi\epsilon)^{\nu/2k}}
\Gamma_i(t) \, c^\dagger \,
e^{{i}\sqrt{{\nu}/{k}}\,\tilde\varphi_i(0)} + h.c. \label{six}
\end{gather}
The fields, $\tilde \vartheta_{R/L}$ and $\tilde \varphi_{R/L}$
satisfy the commutation relations
\begin{gather}
[\tilde \vartheta_{R/L}(x), \tilde \vartheta_{R/L}(x')]=[\tilde
\varphi_{R/L}(x), \tilde \varphi_{R/L}(x')] = 0 \cr [\tilde
\varphi_{R/L}(x), \tilde \vartheta_{R/L}(x')] =\pm i \frac{\pi}{2}
\,{\mathsf{Sgn}}(x-x') \label{seven}
\end{gather}
When $\nu_{eff}={{\nu}/{k}}$ is tuned to unity by tuning $k$,
Eqs.~\eqref{five}-\eqref{seven} together with $\hm_{ad}$ define a
Hamiltonian which is identical to the bosonized version of the
free fermionic Hamiltonian corresponding to the case of $\nu=1$
quantum Hall state describing tunnelling of fermions from freely
propagating left moving edge to a freely propagating right moving
edge via an anti-dot level. The complete the mapping we define the
new free chiral bosonic fields $\tilde\phi_{R}$ and
$\tilde\phi_{L}$ corresponding to the $\nu_{eff}=1$ theory which
are related to $\tilde \varphi_{R/L}(x)$ and $\tilde
\vartheta_{R/L}(x)$ by the following relation
\begin{gather}
\tilde \varphi_{R/L}(x)=\frac{\tilde \phi_{R/L}(x) + \tilde
\phi_{R/L}(-x)}{2}\non\\
 \quad \tilde
\vartheta_{R/L}(x)=\frac{\tilde \phi_{R/L}(x) - \tilde
\phi_{R/L}(-x)}{2} \nonumber
\end{gather}
Such that $[\tilde \phi_{R/L}(x),\tilde \phi_{R/L}(x')]=\pm
i\pi{\mathsf{Sgn}}(x-x') $. It is crucial to note that, in this
new free fermionic theory, every time a fermion tunnels in or out
of the right (left) edge into the anti-dot, it corresponds a net
transfer of charge which is $\nu e$ and not $e$. Once this mapping
to the free fermion theory is established, it is straight-forward
to calculate pumped charge in the adiabatic limit using the
Brouwer's formula~\cite{brouwer}. This is pursued in what follows.

\section{Pumping for the case of $k = \nu$}
Given the mapping established in the previous section, we first
write down the free fermionic Hamiltonian in terms of the Dirac
fields as follows
\begin{gather}
H=-iv_F \int
dx[\psi^{\dag}_R\partial_x\psi_R-\psi^{\dag}_L\partial_x\psi_L] +
\eps c^{\dag}c + \cr
[\Gamma_L(t)c^{\dag}\psi_{L}(0)+\Gamma_R(t)c^{\dag}\psi_{R}(0)+h.c.]
\nonumber
\end{gather}
Here the Fermi velocity, $v_F=v_0 \sqrt{1-\lambda^2}$. Also the
fields $\psi_{R/L}$ is related to the bosonic fields $\tilde
\phi_{R/L}$ by the relation,~$\psi_{R/L}(x)=1/(2 \pi
\epsilon)^{1/2}~e^{i\tilde\phi(x)_{R/L}}$. Here we have dropped
Klein factors as they are not important for our case of single
resonant level. We assume that $\Gamma_{L/R}(t)$ vary periodically
with a period $\tau$, which is much larger then all other time
scales in the problem, so that we are in the adiabatic limit and
hence we can work with the time-frozen Hamiltonian to find the
instantaneous (adiabatic) eigenstates of the Hamiltonian. The
Heisenberg equations of motion
 for the three fields $\psi_R,\psi_L,c$ are
\begin{gather}
\label{eoms}
-i\partial_t \psi_R=iv\partial_x\psi_R+\Gamma_R^\star
c\delta(x)\cr
-i\partial_t\psi_L=-iv\partial_x\psi_L+\Gamma_L^\star
c\delta(x)\cr i\partial_tc=\eps c+
\Gamma_R^{}\psi_R(0)+\Gamma_L^{} \psi_L(0)
\end{gather}
Since for $x\neq0$, The above equations (Eq.~\eqref{eoms}) are
just free particle equations of motion, hence we invoke plane wave
solutions to obtain the scattering matrix as given below. For
algebraic simplification, we define $c=\tilde{c}e^{-i\eps t}$.
\begin{gather}
\psi_R(x,t)=\frac{1}{\sqrt{L}}\sum_{\omega}e^{i\omega(x/v-t)}\left\{
                                                       \begin{array}{ll}
                                                         a_{R,\omega}, & x<0\hbox{;} \\
                                                         b_{R,\omega}, & x>0\hbox{.}
                                                       \end{array}
                                                     \right.\cr
\psi_L(x,t)=\frac{1}{\sqrt{L}}\sum_{\omega}e^{i\omega(x/v+t)}\left\{
                                                       \begin{array}{ll}
                                                         a_{L,\omega}, & x>0\hbox{;} \\
                                                         b_{L,\omega}, & x<0\hbox{.}
                                                       \end{array}
                                                     \right.
                                                     \cr = \frac{1}{\sqrt{L}}\sum_{\omega}e^{i\omega(x/v-t)}\left\{
                                                       \begin{array}{ll}
                                                         a_{L,-\omega}, & x<0\hbox{;} \\
                                                         b_{L,-\omega}, & x>0\hbox{.}
                                                       \end{array}
                                                     \right.\cr
\tilde{c}=\sum_{\omega}e^{-i\omega t}c_{\omega} \nonumber
\end{gather}
Plugging these solutions into \eqref{eoms} we obtain
\begin{gather}
b_{R,\omega}=\frac{|\Gamma_L|^2-|\Gamma_R|^2-iv(\eps-\omega)}
{|\Gamma_R|^2+|\Gamma_L|^2-iv(\eps-\omega)}~a_{R,\omega} \cr -
\frac{2\Gamma_R^\star\Gamma_L}
{|\Gamma_R|^2+|\Gamma_L|^2-iv(\eps-\omega)}~a_{L,-\omega}\cr
b_{L,-\omega}=-\frac{2\Gamma_R\Gamma_L^\star}
{|\Gamma_R|^2+|\Gamma_L|^2-iv(\eps-\omega)}~a_{R,\omega}
\cr+\frac{|\Gamma_R|^2-|\Gamma_L|^2-iv(\eps-\omega)}
{|\Gamma_R|^2+|\Gamma_L|^2-iv(\eps-\omega)}~a_{L,-\omega}
\nonumber
\end{gather}
which gives us the scattering matrix
\begin{gather}
  \begin{vmatrix}
    ~b_{R,\omega}~ \\
    ~b_{L,-\omega}~ \\
    \end{vmatrix}
= S~
 \begin{vmatrix}
                   ~a_{R,\omega}~ \\
                   ~a_{L,-\omega}~ \\
 \end{vmatrix}
\nonumber
\end{gather}
Since we are only interested in the limit, $\omega \to 0$ of the
problem, henceforth we shall only deal with $S$-matrix at $\omega
= 0$. Following~\cite{aleiner} we express the scattering matrix as
\begin{gather}
S=e^{i\theta}~
\begin{vmatrix}
               ~(1-G)^{1/2}e^{-i\beta} & iG^{1/2}~ \\
               ~iG^{1/2} & (1-G)^{1/2}e^{i\beta}~ \\
             \end{vmatrix}
\nonumber
\end{gather}
Here $G$ is the dimension-less instantaneous conductance and is
given by
\begin{gather}
\label{fifteen}
G(t)=-\left|\frac{2\Gamma_{R}(t)\Gamma_{L}(t)}
{\Gamma_{R}(t)^2+\Gamma_{L}(t)^2-iv\eps}\right|^2
\end{gather}
Without loss of generality,  we have assumed that $\Gamma_{R/L}$
are real quantities in Eq.~\eqref{fifteen}.
 Upon using the Brouwer's formula, the pumped charge
can be be straight-forwardly obtained as
\begin{gather}
{\cal Q}=\frac{e^\star}{2\pi}\int_0^{\tau}
dt\left[\frac{d\lambda(t)}{dt}-G(t)\frac{d\beta(t)}{dt}\right]
\label{pumpedch}
\end{gather}
where the integral is taken along the closed contour in the
parameter space of $\Gamma_{R}-\Gamma_{L}$,  during one pumping
cycle and $\lambda=\theta-\beta$ is a reflection amplitude phase.
The first term gives a quantized contribution to pumped charge in
units of $e^\star={\nu}e$. It is topological in nature and hence
does not depend on the details of the contour. This is because
$\lambda(\tau)=\lambda(0)+ 2n\pi$ where $n$ is an integer. The
second term is the one which destroys quantization as it is
directly proportional to the conductance.

From the above equation (Eq.~\eqref{fifteen}) we note that if the
resonant level is tuned such that its position in energy space is
infinitesimally close to the Fermi energy of the leads, then the
line corresponding to $\Gamma_{R}=\Gamma_{L}$ in the
$\Gamma_{R}-\Gamma_{L}$ plane corresponds to a line of perfect
resonances \ie,~ $G=1$. To obtain quantized pumped
charge it is essential that \\
{\textsl{(a)}} The topological contribution to pumped charge is
non-zero. This is achieved by choosing the pumping contour to be
such that, it encloses a finite portion of the line of resonances.
This is because every time the pumping contour crosses the line of
resonances, the phase of the reflection amplitude discontinuously
changes by a factor of $\pi$. As the pumping contour will cross
the line of resonances at least at two points for the case of
simplest possible closed curve in the parameter plane, the total
change in reflection phase $\theta$ over one complete pumping
cycle amounts to a total change of $2\pi$ resulting in pumped
charge of amount ${\nu}e$.
\\
{\textsl{(b)}} It is also important to note that, in the
expression for the pumped charge there is part which is
proportional to conductance (dissipative part) which will destroy
the quantization until and unless the pumping contour is such that
the value of conductance $G(t)$ remains vanishingly small on the
pumping contour. This can be achieved by tuning the couplings
$\Gamma_{R/L}$ to be very small such that the resonance become
sharp leading to vanishing of $G(t)$ in most part of the pumping
contour, as is evident from Eq.~\eqref{fifteen}. Now the only
hurdle which still remains is related to the fact that the
dissipative part picks up considerable contribution
when the pumping contour cuts the line of resonances.\\
This hurdle is a consequence of the fact that we have  used an
over-simplified model for the resonant level. Generically, the
position of the resonant level which corresponds to quasi-bound
state in the anti-dot is not independent of $\Gamma_{R/L}$, but is
expected to be a smooth function of $\Gamma_{R/L}$~\cite{levinson}.
Hence the line of perfect resonance at $\Gamma_{R}=\Gamma_{L}$ for a
given Fermi energy in the lead will get restricted and will shrink
to a set of points in the parameter space which can be enclosed
completely within an appropriately chosen pumping contour leading to
almost perfect quantization of pumped charge. We say ``almost"
because there will always be some  small yet finite contribution
coming from  the tail of the resonance which will fall on the
pumping contour. It is worth noticing that even when the pumping
contour does not cross any line of resonance and only encloses a
point of perfect resonance in the pumping parameter space, the phase
of the reflection amplitude changes in multiples of $2\pi$ over a
period $\tau$, and is not a periodic function of time. This is
because the resonance corresponds to zero refection ($r=0$) and
hence $\lambda={{\cal Im}}\,[\log r]$ goes through a branch cut when
the contour encloses the resonance which ultimately leads to the
non-periodicity of reflection phase as a function of time.
\section{Pumping for the case of $k \neq \nu$}
For the case of $k {\neq} 1/3$, we study the problem
perturbatively in the parameter ${\delta} k$ which represents a
small deviation of $k$ around the point, $k=1/3$. Since this point
corresponds to free fermionic theory, the case of $k = ({1}/{3})
{\pm} {\delta} k $ maps onto the problem of weakly interacting
spin-less fermion (non-chiral Luttinger liquid) whose Luttinger
parameter is given by $(k{\pm}~{\delta}k)/\nu\,$ which is
$1{\pm}({\delta}k/\nu)$ when $k=\nu$. The plus sign corresponds to
the case of attractive fermions while the minus sign corresponds
to repulsive fermions. Also we assume that ${\delta}k << {\nu}$.
For the case of weakly interacting fermions in \odd, it is
possible to calculate transport through a localized scalar
impurity~\cite{matveev} or a resonant level~\cite{glazman} by
calculating corrections to the free fermion scattering matrix
element representing the quantum scatterer, perturbatively in the
interaction strength followed by a ``poor-man's scaling" approach
which eventually gives an \rg equation for the $S$-matrix element
itself. As an alternative to bosonization, we adopt the above
mentioned approach to first calculate the $S$-matrix elements at a
given energy scale corresponding to the time-frozen Hamiltonian
for the problem at hand. We then solve the \rg equations for the
$S$-matrix to obtain the energy scale dependence of these
elements. Then using the Brouwer's formula, we calculate the
pumped charge in the adiabatic limit for the case $k =
\nu{\pm}{\delta}k$. The \rg equation for transmission amplitude of
the fermions through the resonant level, when the level is tuned
infinitesimally close to the Fermi energy in the lead is given
by~\cite{glazman}
\bea {d{t_{R/L}^{\pm}}\over d\,{\ln({{\Delta E}/\tilde
\epsilon})}} &=&
{\pm}\left({{\delta}k\over\nu}\right)\,\left[{t_{R/L}^\pm}\,
\left\{1-\left|{t_{R/L}^\pm}\right|^2\right\} \right]
\label{transRg}\eea
\bea {dr_{R/L}^{\pm}\over d\,{\ln({{\Delta E}/\tilde
\epsilon})}}&=& \mp \left({{\delta}k\over {\nu}}\right)
\Big[-r_{R/L}^{\pm} + r_{R/L}^{\pm} \vert r_{R/L}^{\pm} \vert^2
\non\\ && +~ {t_{R/L}^{\pm}} {t_{L/R}^{\pm}} {r_{L/R}^{\pm}}\Big]
\eea
Here, ${t_{R/L}^{\pm}}$ represents the transmission amplitude from
the right (left) moving edge to the left (right) moving edge via the
resonant level when the effective Luttinger liquid parameter is
tuned to $1{\pm}{\delta{k}/{\nu}}$. Similarly, ${r_{R/L}^{\pm}}$
stands for the reflection amplitude in the right (left) moving edge
respectively. And $\tilde \epsilon$ is the energy of the {{fermion}}
measured form the Fermi-level. The ultra-violet cut-off scale in our
case is set by the average level spacing of the levels in the
anti-dot ($\Delta E$). Now, given the fact that we are at zero bias
and the pumping frequency is such that it is smaller then all the
relevant energy scale ($\eps,\Delta E$) in the problem to remain in
the adiabatic limit, hence the only energy scale which acts as the
low energy cutoff is energy scale set by temperature and we must
stop the \rg flow at energy corresponding to $|\epsilon|\simeq k_B
T$. So, the solutions of the \rg equation will provide us with the
temperature dependence of the transmission and reflection
amplitudes. Note that the small parameter corresponding to the
interaction strength (say $\alpha$) in terms of which the fermionic
perturbation theory is developed is connected to the Luttinger
parameter (say $K$) of the fermions via
$K=\sqrt{(1-{\alpha})/(1+{\alpha})}$~\cite{fisher}. As we are only
interested in the weak interaction limit,~\ie~the limit of $\alpha
\to 0$ we expand $K$ to first order in $\alpha$ to get
$K=1-{\alpha}$. For fermions with attractive interaction, we have $K
> 1$ and $\alpha$ is negative while for the fermions with repulsive
interaction, $K < 1$ and $\alpha$ is positive. From
Eq.~\eqref{transRg}, it is easy to see that the transmission
amplitude for the case of attractive fermions $(K=1+{\delta}k/\nu
\Rightarrow \alpha=-\delta k/\nu)$ grows under \rg flow. Hence the
\rg flow will destroy the quasi-particle resonant level in the \ir
limit and quantization of pumped charge in units of ${\nu}e$ cannot
be achieved in this case. On the contrary, for the case of repulsive
fermions $(K=1-{\delta}k/\nu \Rightarrow \alpha=\delta k/\nu)$,
Eq.~\eqref{transRg} suggests that the transmission amplitude flows
to zero under \rg  whenever the transmission through the level is
different then unity when the \ir limit is taken. Hence in this case
the \rg flow leads to an extremely sharp resonance resulting in
quantization of pumped charge in units of ${\nu}e$ in the \ir limit.
 {As a matter of fact, similar results regarding
 quantization of pumped charge were also obtained in
 Ref.~\cite{chamon,prashant} for the case of $K<1$ and $K>1$.}

{Now,} after obtaining the temperature dependence of the $S$-matrix
elements for the level, we can obtain the temperature dependence of
the pumped charge at a given temperature using Eq.~\eqref{pumpedch}.
It is worth noticing that Eq.~\eqref{transRg} leads to \rg flow
equation for both the transmission $(|t_{R/L}^{\pm}|)$ and
reflection amplitude ($|r_{R/L}^{\pm}|$) and the associated phases
$\lambda=\theta-\beta,{\theta}$ respectively. Without loss of
generality, we can choose to calculate the temperature dependence of
the $S$-matrix element anywhere on the pumping contour in the space
of pumping parameters ({\ie}~$\Gamma_R - \Gamma_L$ ). To avoid
unnecessary complications arising due to the \rg flow of phases
associated with $S$-matrix elements ($\lambda,\theta$), we choose to
calculate the solution for the \rg equation for the case of
symmetric barrier. This symmetry leads to vanishing of the \rg flow
of the phase hence simplifying the calculation considerably. In
doing so we have assumed that, the pumped charge at different
temperature had been obtained by cooling the system when the
coupling of the resonant level is tuned to be symmetric with respect
to the left moving edge and the right moving edge \ie,~at those
points on the pumping contour where $\Gamma_1=\Gamma_2$ line cuts
the pumping contour. The \rg equation for the transmission
 amplitude for the symmetric case hence reduces to
\bea {d|{t_{R/L}^{\pm}}|\over dl} &=&
{\pm}\left({{\delta}k\over\nu}\right)\,\left[~|{t_{R/L}^\pm}|\,
\left\{1-|{t_{R/L}^\pm}|^2\right\} ~\right] \label{transRg1} \eea
We now integrate the \rg equation for obtaining the power-law
dependence for the dimension-less
conductance($(|t_{R/L}^{\pm}|^2)$) as a function of temperature as
\bea G^{\pm}(T) &=& {T_0^{\pm}} \frac{\left(\frac{\Delta E}{k_B
T}\right)^{\pm 2(\delta k/\nu)}}{\left[R^{\pm}_{0} +
T^{\pm}_{0}\left\{ \left(\frac{\Delta E}{k_B T}\right)^{\pm
2(\delta k/\nu)}\right\}\right]} \label{ten} \eea
Here ${T_0^{\pm}}$ and ${R_0^{\pm}}$ are the values of $T^{\pm}$
and $R^{\pm}$ at $L_T=d$. Hence using Eqs.~\eqref{pumpedch} and
\eqref{ten}, we obtain the pumped charge as a function of
temperature which is given by
\bea {\cal Q}^\pm &=& {\cal Q}_{\mathrm {int}}- \left(\frac{\Delta
E}{k_B T}\right)^{\pm 2 (\delta k/\nu)}\
\int_0^\tau dt~ I^{\pm}(t) \label{qq}
\\
{\mathrm {where}}~\nonumber \\
I^{\pm}(t) &=& \frac{e^\star}{2\pi}
 \frac{G^{\pm}_0 \left[ \dot \beta \right]}{1 + G^{\pm}_0\left[-1 +
\left(\frac{\Delta E}{k_B T}\right)^{\pm 2 \delta k/\nu} \right]}
\nonumber  \eea Here $I^{\pm}(t)$ represents the instantaneous
dissipative current which spoils the quantization of pumped charge.
From Eq.~\eqref{qq}, it is easy to see that  for the case of $k >
1/3$ \ie,~ for ${\cal Q}^+$, quantization of the pumped charge
becomes worse as we go to lower temperatures, which is consistent
with the observation that for $k=1$ the quantization is in integer
units of $e$, whereas for the case of $k < 1/3$ \ie,~ for ${\cal
Q}^-$, the quantization of the pumped charge is achieved as we go to
lower temperatures limit (\ir limit).

\section{Conclusions and Discussion}
In conclusion, in this letter we obtain an exact expression for the
pumped charge in the adiabatic limit for a specific strength of the
inter-edge interaction, namely for $k = 1/3$. We also obtain an
expression for pumped charge for any small deviation around the
value of $k=1/3$ perturbatively in the small parameter which is
taken to be the deviation of $k$ from its value of one-thirds and
show that for $k\leq1/3$ the charge pumped in a cycle is quantized
in units of $\nu e$. We also propose a experimental set-up in which
our prediction regarding quantization of pumped charge in units of
${\nu}e$ can be verified. {In the end, it is important to emphasize
that for temperatures $T<T_{L/2}$ ($T_{L/2}$ being the temperature
corresponding to the distance $L/2$, the length of each line
junction), there will be correlation to the power law dependance of
the conductance due to junction between the edge of the line
junction and the freely propagating edge connected to the contacts
(Fig.~\ref{geometry}(b)). This would imply that the results quoted
in Eq.~\eqref{qq} should get appropriately modified for temperatures
$T<T_{L/2}$.  To avoid this complication one can fabricate ohmic
contacts right at the end of the two line junctions rather than
having it far from the line junction as in Fig.~\ref{geometry}(b)
and hence for $T<T_{L/2}$, the temperature power laws in
Eq.~\eqref{qq} are to be replaced by  length power laws.}

\section{Acknowledgements}
We thank Yuval Gefen for many stimulating discussions on the
physics of edge state, comments on the manuscript and
encouragement. SD gratefully acknowledges Sumathi Rao, Efrat
Shimshoni and Izhar Neder for useful discussions. We acknowledge
financial support from the Feinberg Fellowship at Weizmann
Institute of Science, Israel, the US$-$Israel BSF, the Albert
Einstein Minerva Center for Theoretical Physics and a grant from
the Minerva Foundation.

\bibliographystyle{eplbib} 
%
\bibliography{myref} 
\end{document}